# The Band Gap of Graphene Is Efficiently Tuned by Monovalent Ions

Guilherme Colherinhas[1], Eudes Eterno Fileti[2] and Vitaly V. Chaban[2]

1) Departamento de Física, CEPAE, Universidade Federal de Goiás, CP.131, 74001-970, Goiânia, GO, Brazil

2) Instituto de Ciência e Tecnologia, Universidade Federal de São Paulo, 12247-014, São José dos Campos, SP, Brazil

**Abstract**. Following recently published study of Prezhdo and coworkers (*JPC Letters, 2014, 5, 4129-4133*), we report a systematic investigation of how monovalent and divalent ions influence valence electronic structure of graphene. Pure density functional theory is employed to compute electronic energy levels. We show that LUMO of an alkali ion ($Li^+$, $Na^+$) fits between HOMO and LUMO of graphene, in such a way tuning the bottom of the conduction band (i.e. band gap). In turn, $Mg^{2+}$ shares its orbitals with graphene. The corresponding binding energy is ca. 4 times higher than in the case of alkali ions. The reported insights provide inspiration for engineering electrical properties of the graphene containing systems.

**Introduction**

Carbonaceous materials offering nanoscale pores are important for a variety of promising applications and technologies.[1-17] Most known examples include sensing and catalytic devices,[16,17] electrochemical tools,[7,9-11,18] and biomedical setups.[12-14] Graphene and its numerous derivatives are currently pursued actively in connection to supercapacitors and lithium-ion batteries.[15,19-26] These applications imply direct interaction of the lithium ions with a carbonaceous material, including intercalation of the ions during the charging cycle. These interactions appear, in many cases, non-intuitive for a chemist, while they strongly depend on external conditions and chemical environment.[10,15,24] The successful intercalation also correlates with the size of the pore (smaller pores perform better) and the presence of dopants on its surface. Binding between nanoscale carbonaceous structures and light ions cannot be described by simple mathematical expressions, originating from general chemical wisdom. The numerical simulation techniques provide an efficient and often reliable tool to obtain necessary physical insights for each system of practical interest.

In a very recent publication in *The Journal of Physical Chemistry Letters*, Prezhdo and coworkers[27] demonstrated that the lowest unoccupied molecular orbital (LUMO) of the lithium ion is sensitive to the distance of the ion from the sidewall of semiconductor single-walled carbon nanotube. This particular orbital also fits between the highest occupied molecular orbital (HOMO) and LUMO of the nanotube. Therefore, vacant electron energy level, which arrives from lithium, shapes a new bottom of the conduction band in this system. The band gap of the whole system can be, to certain extent, tuned by the ion location within an inner cavity of semiconducting nanotube. The above conclusions are based on the pure density functional theory (DFT) with a plane-wave basis set. Periodic boundary conditions coupled with the Brillouin zone sampling were used to represent virtually infinite carbon nanotube, irrespective of the actual supercell size.

In the present work, we extend and generalize studies of Prezhdo and coworkers by employing lithium ($Li^+$), sodium ($Na^+$), and magnesium ($Mg^{2+}$). Instead of carbon nanotube, we consider a graphene-like sheet. First, graphene features – in many aspects – similar to nanotube electronic structure, i.e. delocalized polarizable π-electrons. We refrain from using periodic boundary conditions to avoid methodological problems with the infinite charged system. Convergence of the wave function requires an implementation of the terminating hydrogen atoms. The implemented model is more general than periodic graphene, since it accounts for changeable chemical environment (hydrogen atoms). Note, that the band gap of real graphene is determined by its π-electrons, while terminating hydrogen atoms, with their *s*-electrons, do not participate in the band gap formation. The graphene sheets, which are not large enough, exhibit properties of a conventional finite-gap semiconductor. In the following, we will refer to this model simply as *graphene*, keeping in mind the underlined divergences from a real graphene sheet.

Figures 1-3 depict localization of HOMO and LUMO of graphene, as well as LUMO of lithium. Small distances between graphene and the ion (0.1-0.2 nm) can be understood as a consequence of high external pressure. These cases constitute certain practical interest, as high pressure is able to essentially alter electronic energy levels and orbital localizations. For instance, decrease of a comfortable separation between $Li^+$ and graphene by just 0.2 nm leads to a drastic jump of $Li^+$ LUMO by 2.077 eV upwards. The resulting orbital becomes the 9$^{th}$ molecular orbital above LUMO in the [Li@graphene]$^+$ complex. LUMO in this complex is also localized on graphene. Similar behavior of LUMO of the ion is observed for $Na^+$ and $Mg^{2+}$ under compression.

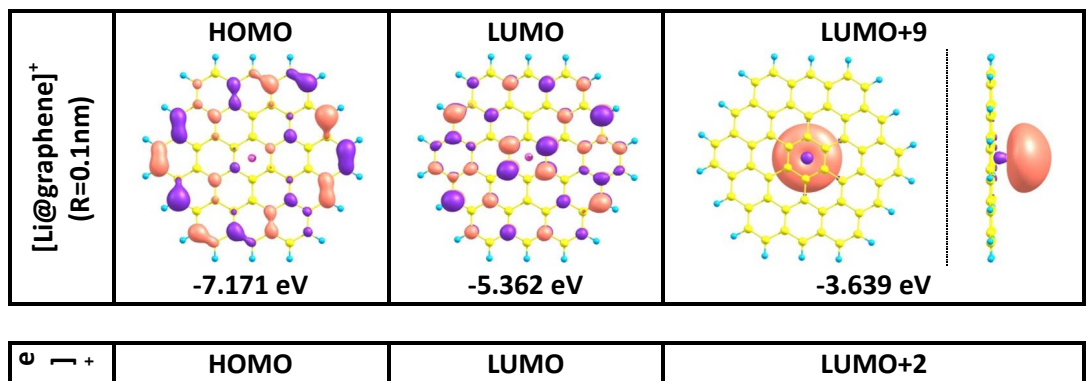

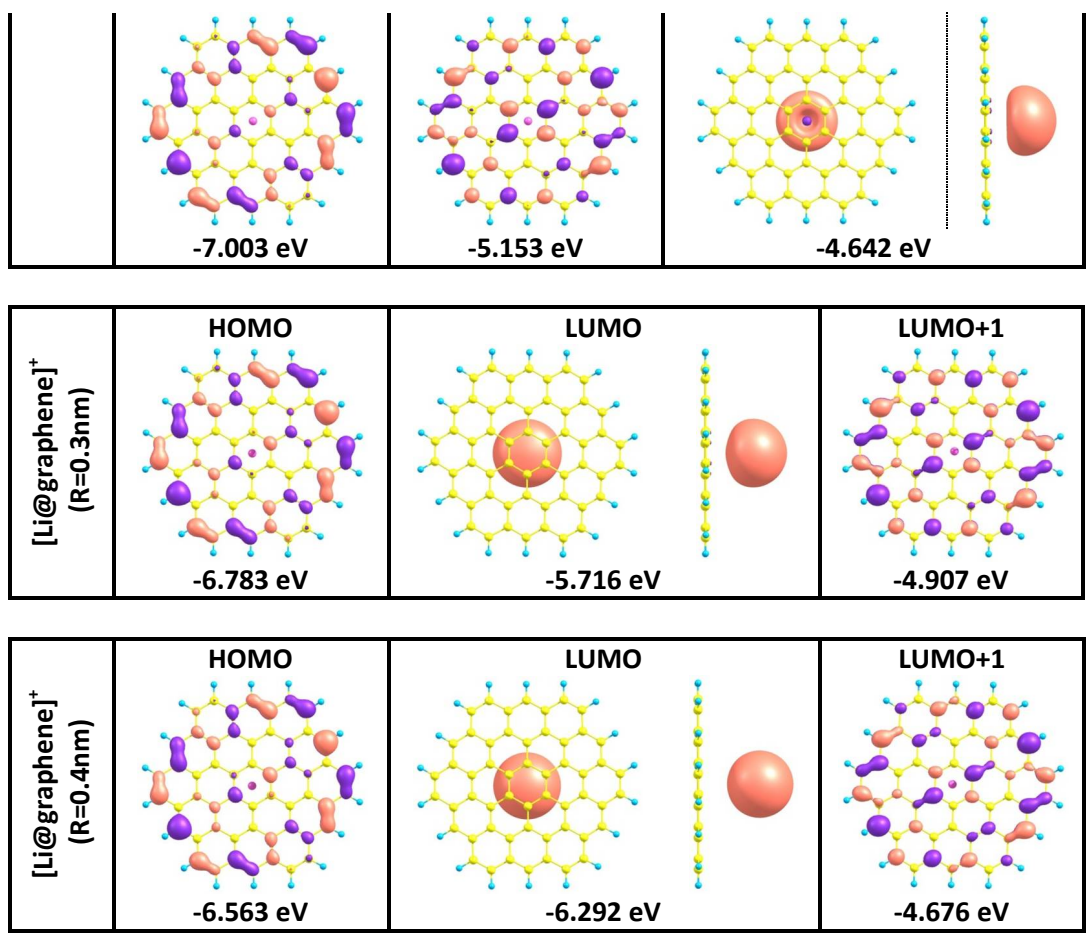

Figure 1. Spatial localization of HOMO and LUMO in a series of [Li@graphene]$^+$ complexes. The depicted molecular orbitals were obtained using BLYP DFT[28] employing the LANL2DZ basis set. Whereas HOMO is routinely localized on graphene, localization of LUMO depends on the distance between graphene and lithium ion.

The cases of [Li@graphene]$^+$ and [Na@graphene]$^+$ are qualitatively same. The energies of LUMOs in both ions fit between HOMO and LUMO of graphene at certain separations (Figure 4). The observed quantitative difference is plausibly explained by the variation of the van der Waals radius in these alkali cations. LUMOs of both cations form a new bottom of the conduction band in the corresponding systems. It should be expected that larger separations correlate with smaller ionic LUMO energies. In turn, valence bands are driven by the electron-electron repulsion in the systems under pressure.

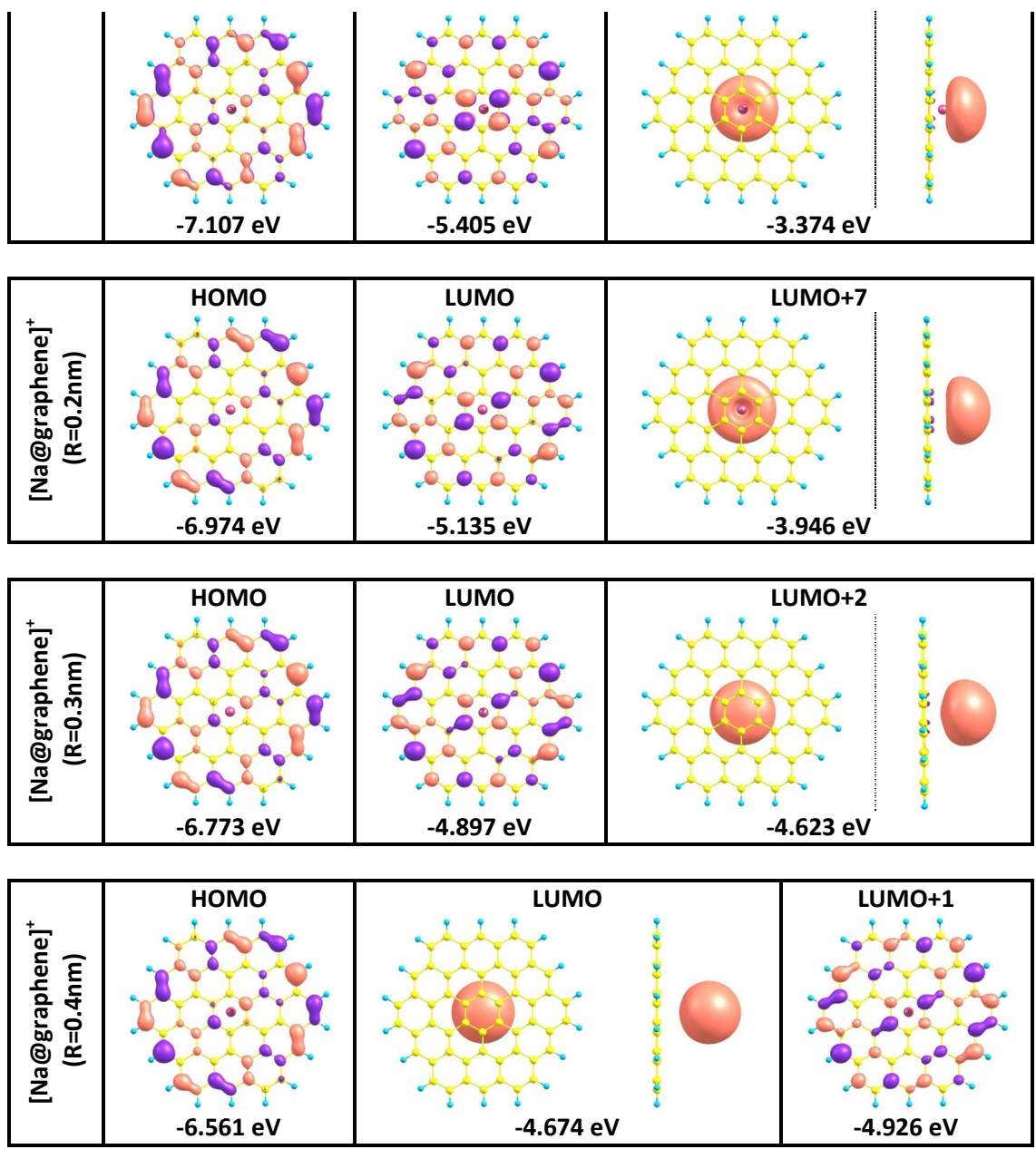

Figure 2. Spatial localization of HOMO and LUMO in a series of [Na@graphene]$^+$ systems. The depicted molecular orbitals were obtained using BLYP DFT[28] employing the LANL2DZ basis set. Whereas HOMO is always localized on graphene, localization of LUMO depends on the distance between graphene and sodium ion.

The divalent ion, Mg$^{2+}$, exhibits a qualitatively different behavior (Figure 3), as compared to alkali ions. Its LUMO does not populate the band gap of graphene. In turn, LUMO of the [Mg@graphene]$^{2+}$ complex is shared by both magnesium atom and graphene sheet. This implies strong binding in the complex and partial charge transfer of the higher-energy electrons from graphene to an electron deficient cation. As a result, LUMO of Mg$^{2+}$ and HOMO of graphene become nearly degenerate (Figure 4). Despite strong binding to the π-electron system, the divalent ion cannot be used to tune the band gap of our graphene sheet.

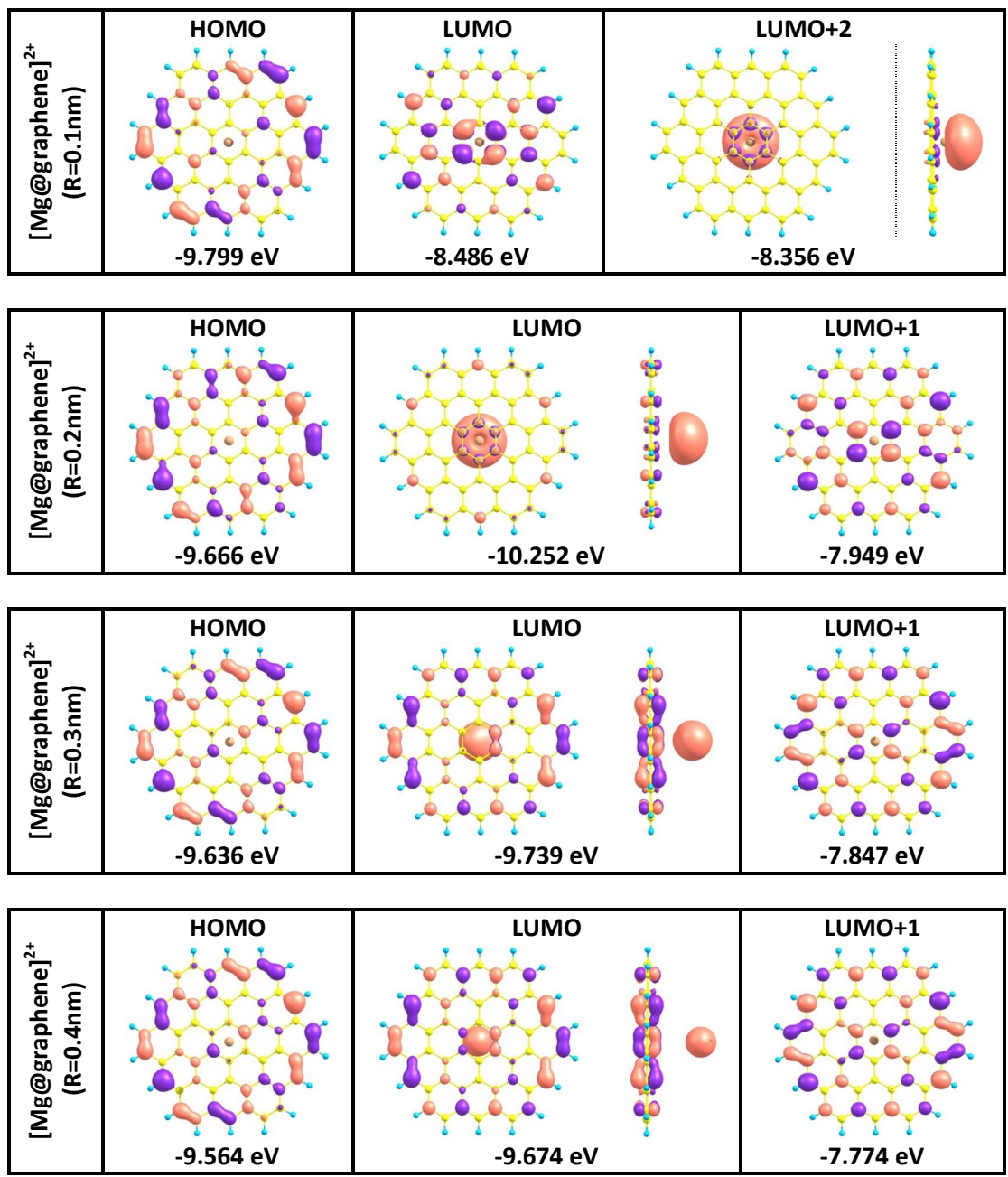

Figure 3. Spatial localization of HOMO and LUMO in a series of [Mg@graphene]$^{2+}$ systems. The depicted molecular orbitals were obtained using BLYP DFT[28] employing the LANL2DZ basis set. Whereas HOMO is always localized on graphene, LUMO is shared between graphene and magnesium ion. In turn, LUMO+1 is localized on graphene, irrespective of the separation distance.

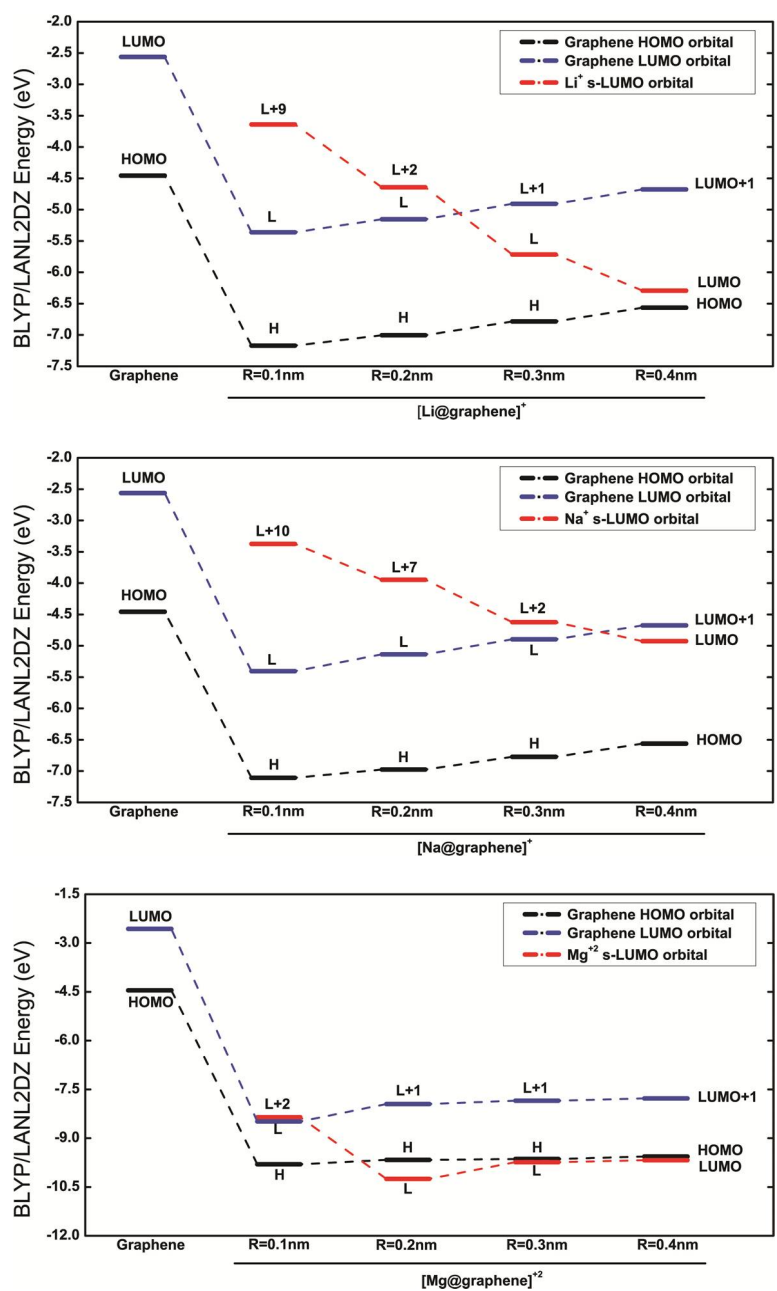

Figure 4. Evolution of valence orbital energies in the [ion@graphene] complexes as a function of separation. The energies of HOMO and LUMO of graphene get remarkably lower after an ion is added to the system.

Interaction energies and basis set superposition errors (BSSE) plotted as a function of separation distance (Figure 5) provide important information regarding (1) chemical or physical nature of binding in the supramolecular complex; (2) relative strength of binding to compare an effect of various ions; (3) position of potential energy minimum. Both $Li^+$ and $Na^+$ exhibit binding energies to graphene of less than 50 kcal mol$^{-1}$. The distances of closest approach under zero external pressure equal to 0.18 nm ($Li^+$) and 0.22 nm ($Na^+$). Compare, the experimentally determined covalent radius of carbon is 0.077 nm, whereas the covalent radii of lithium and sodium are 0.134 and 0.154 nm, respectively. Note that distances in Figure 5 correspond to the surface-nucleus distances rather than to nucleus-nucleus ones. Therefore, 0.18 vs. 0.21 nm ($Li^+$) and 0.22 vs. 0.23 nm ($Na^+$) correlate appropriately. If the interaction between the alkali ions and the graphene sheet were limited to van der Waals binding, the computed separation distances would have been significantly larger.

Binding of $Mg^{2+}$ to graphene is about four times stronger than that of alkali ions. This amount of binding energy is comparable to that of many single covalent bonds. Partial charge transfer (Figure 6) is observed from graphene to magnesium. Both species become partially positively charged. Coulomb repulsion component is added to the total complexation energy. This effect is responsible for an unusual tail (0.3-0.5 nm) of the binding energy curve. As the species get farther from one another, Coulombic repulsion contribution decreases. In addition, amount of charge transfer decreases upon separation (Figure 6).

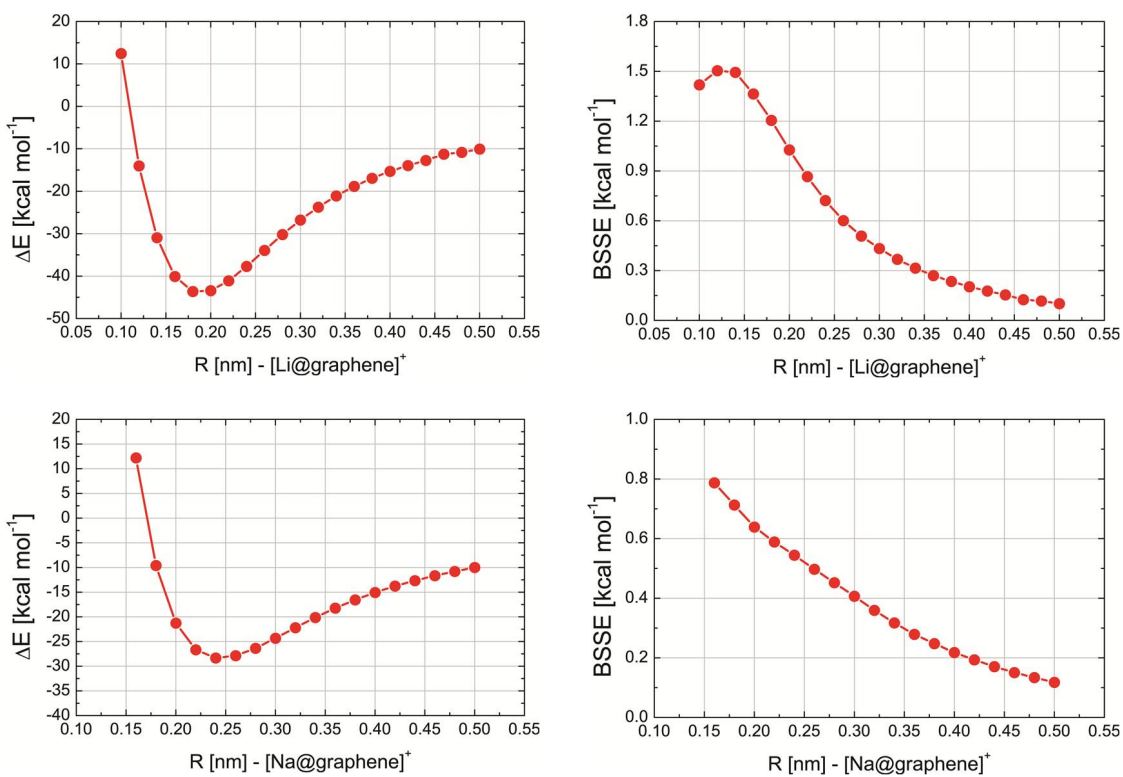

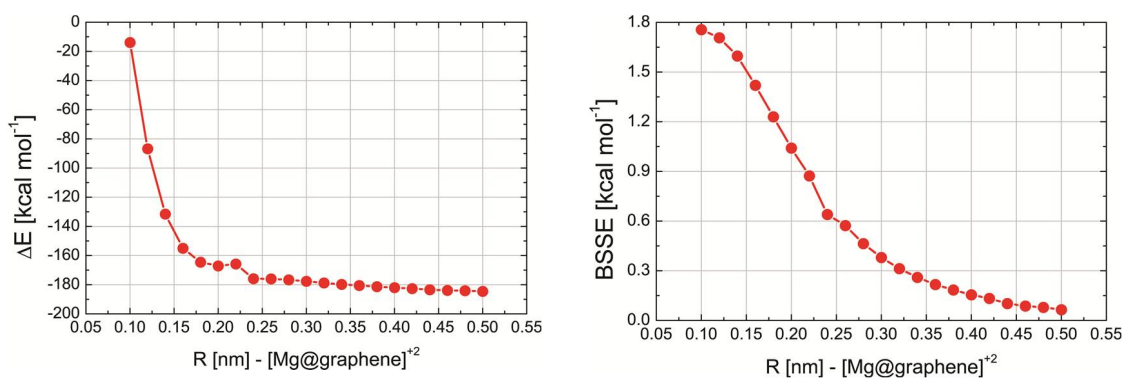

Figure 5. Binding energies maintaining an [ion@graphene] (Li$^+$, Na$^+$, Mg$^{2+}$) complex as a whole. The interaction is essential in all cases significantly exceeding van der Waals forces. The binding strength depends on the size and the charge of an ion. Basis set superposition error (BSSE) corrections are provided for estimation being in all cases inferior to binding energies. Small ratio of BSSE to binding energy constitutes another evidence of the non-VDW nature of the considered complexes. In VDW complexes, BSSE may amount to ca. 50% of binding energy.

Figure 6 evidences that charges on the lithium, sodium, and magnesium ions deviate from +1e (+2e) appreciably, whereas a particular amount depends on their closeness to the graphene surface. The smaller is the ion, the larger portion of electronic charge it obtains from graphene at the covalent-distance separations. Heavy alterations of electronic population upon complexation are in concordance with Figures 1-5, which suggest a strong polarizing action of each cation. Due to delocalization (high potential energy) of carbon's electrons, graphene readily responds to such a polarizing action.

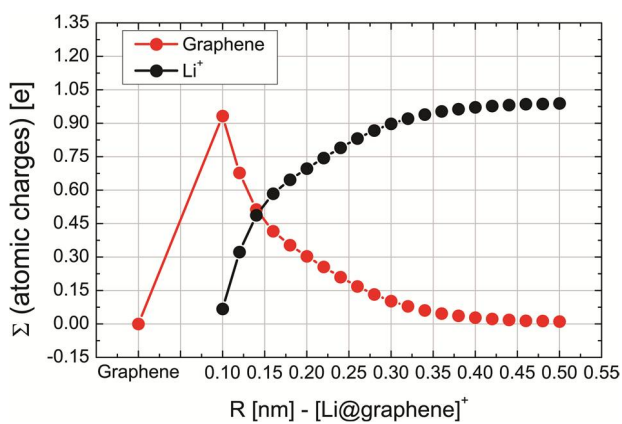

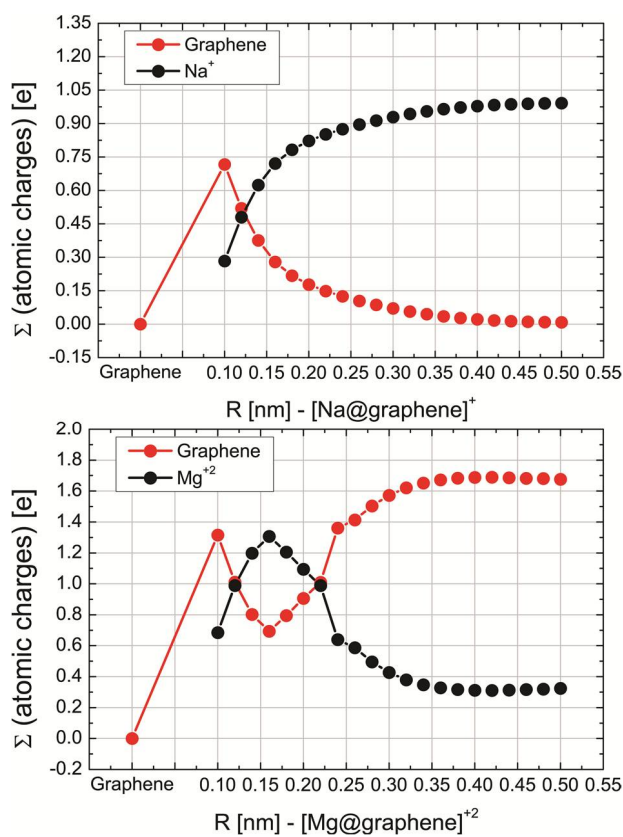

Figure 6. Mulliken atomic charges on graphene and the ion as a function of separation distance. Note, that isolated graphene is electrostatically neutral, while an isolated ion is charged (+1e, +2e). The case of $Mg^{2+}$ is qualitatively different from monovalent cations ($Li^+$, $Na^+$). As separation distance approaches the sum of van der Waals radii of carbon and magnesium atoms, most of positive charge is transferred to graphene. To rephrase, valence electrons of graphene are shifted to Mg (see localization of LUMO, Figure 3). Despite both graphene and the ion become positively charged (electrostatic repulsion), the corresponding complexation energy is large (~180 kcal mol$^{-1}$, Figure 5).

Polarization of graphene by an ion can be conveniently quantified based on the alteration of partial atomic charges of pristine graphene. In principle, any sort of atomic charges can be applied, since all of them reflect a non-uniform distribution of electronic density among atomic nuclei. We introduce an integral measure of polarization using the sum of squared atomic charges localized on every atom of the graphene sheet. It is a remarkable observation that $f(q_i)$ is highly sensitive to the ion presence and the ion nature, but is relatively insensitive to the ion distance from the graphene surface. Note, however, that $f(q_i)$ is, in most systems with ions, smaller than $f(q_i)$ in the pure graphene system. Terminating hydrogen atoms provide a principal contribution to $f(q_i)$, although they are inert in relation to band gap tuning. Therefore, $f(q_i)$ is systematically larger than zero. Adsorbed ions, with an exception of $Li^+$ at 0.3 nm, decrease $f(q_i)$. Each ion breaks the symmetry of the pristine graphene sheet. In all cases, ion induced changes appear much larger than numerical uncertainty in the corresponding calculations.

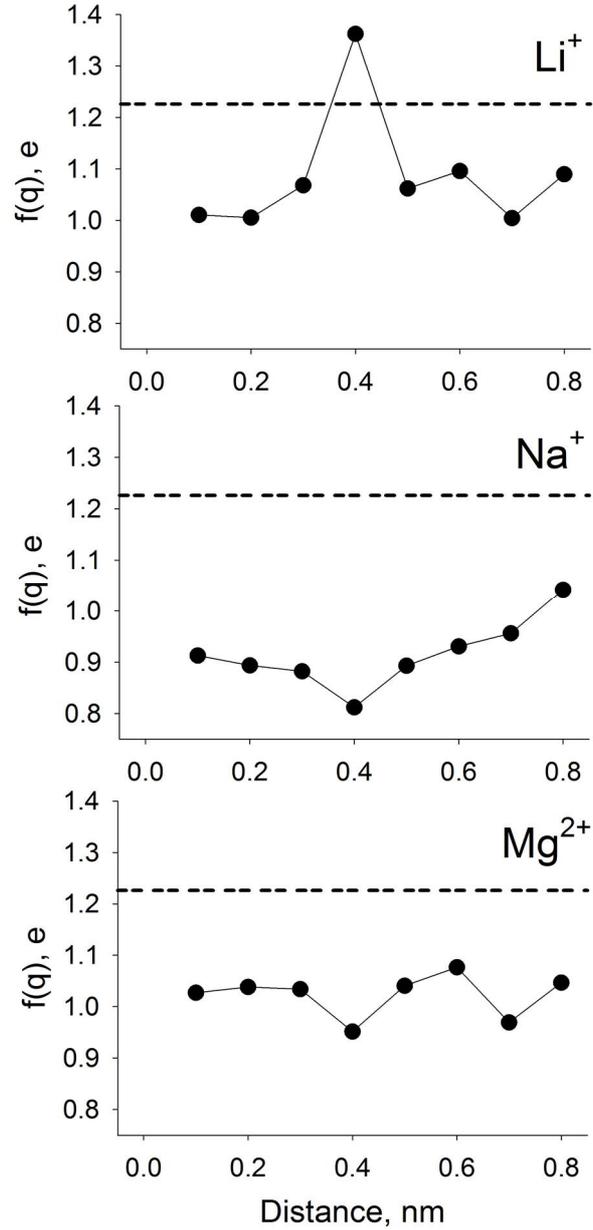

Figure 7. Polarizing action of an ion imposed on the graphene-like sheet, expressed as $f(q_i) = \sqrt{\sum_i^N q_i^2}$. The dashed horizontal line corresponds to $f(q_i)$ of the isolated sheet. Recall, that graphene model, which we use in the present study, is terminated by hydrogen atoms. Consequently, $f(q_i)$ is not close to zero, as it would be in the case of fullerenes or other carbon-only chemical formations.

Using pure density functional theory with the BLYP exchange-correlation functional and LANL2DZ basis set, we investigate the changes in valence and conduction energy levels of graphene and $Li^+$, $Na^+$, $Mg^{2+}$ due to their strong polarizing action. The energies of both HOMO and LUMO of graphene decrease significantly when paired with ions. In turn, LUMO energies of $Li^+$ and $Na^+$ increase as the ions approach graphene surface. The stepwise alteration of these energies allows for simple band gap engineering. For instance, small separations can be achieved at high pressures, while larger separations may be a result of solvation by a proper solvent or coordination by a suitable anion. It should be kept in mind that solvation and neutralization (addition of a counter-ion) are band gap changing events per se. The possibility to use alkali ions to tune the band gaps of carbonaceous aromatic systems is in line with recent result of Prezhdo and coworkers.[27] The present work concentrates on the simplified charged systems. The main result suggests a principal possibility to tune the band gaps of the semiconducting graphene-like systems (and generally an arbitrary carbonaceous aromatic supramolecular system) by just changing its distance to an ion. On another hand, such a strong alteration of the band gap reflects strong non-additivity of the considered interactions. The case of $Mg^{2+}$ is qualitatively different from that of monovalent ions. Initial HOMO and graphene and LUMO of $Mg^{2+}$ are shared in their complex, while the corresponding energy levels nearly degenerate. Therefore, divalent cations, unlike monovalent ones, do not provide an opportunity to tune band gaps. Divalent cations can be efficiently used to induce positive charges on graphene.

**Methodology**

All the reported numerical results were obtained using pure density functional theory. The BLYP exchange-correlation functional[28] was used to calculate electronic energy levels, hybridized orbitals, partial atomic charges, and interaction energies. BLYP is a well-established, reliable functional in the generalized gradient approximation. The wave function was expanded using the effective-core LANL2DZ basis set developed by Dunning and coworkers. This basis provides a reasonable tradeoff between accuracy and computational expense, suitable for MD simulations with average-size systems. The wave function convergence criterion was set to $10^{-8}$ Hartree for all calculations. More accurate calculations would require a larger basis set and a hybrid DFT functional. Pure DFT functionals, such as BLYP,[28] tend to overestimate electronic polarization by favoring delocalized electrons. Note, however, that usage of the moderate size basis set, such as LANL2DZ, counteracts this tendency.

Graphene containing quantum chemical systems often feature difficult self-consistent field (SCF) convergence cases. It occurs due to a large number of higher-energy, delocalized, valence π-electrons on the graphene surface. Additionally, relatively large systems converge slower than smaller systems of the same chemical identity. To guarantee SCF convergence, we employed a quadratically convergent (QC) SCF procedure.[29,30] This procedure consists of linear searches when SCF convergence is far and Newton-Raphson steps when SCF convergence is close. QC SCF is significantly slower[29,30] than regular SCF with DIIS extrapolation, but it is more reliable. Furthermore, the selected SCF procedure accounted for integral symmetry by replicating the corresponding integrals using symmetry operations.

The molecular orbitals (Figures 1-3) and energy levels (Figure 4) were computed for optimized geometries of the [Li$^+$ (Na$^+$, Mg$^{2+}$)@graphene] complexes, whereas the ion was uniformly pulled away from 0.1 to 0.4 nm with a step size of 0.1 nm. In the case of binding energy calculations (Figure 5), an ion was pulled away from 0.1 to 0.5 nm with a step of 0.02 nm. Basis set superposition error was computed for all considered geometries. The depicted energies have BSSE deducted. BSSE decreases as ions get farther from the graphene surface. The electron delocalization (Figure 6) is characterized in terms of Mulliken charges. The polarizing action of the ions is given by the following function, $f(q_i) = \sqrt{\sum_i^N q_i^2}$, where $q_i$ enumerates partial point charges on every atom belonging to graphene. The charges derived from electrostatic potentials on the complex surface were used in the above formula. The atom spheres were defined according to the CHELPG scheme.[31,32]

The electronic structure computations were performed using GAUSSIAN 09, revision D (*www.gaussian.com*). The subsequent analysis was performed using simple utilities developed by the authors.

**Acknowledgments**

E.E.F. thanks Brazilian agencies FAPESP and CNPq for support. V.V.C. acknowledges research grant from CAPES (Coordenação de Aperfeiçoamento de Pessoal de Nível Superior, Brasil) under "Science Without Borders" program.

**AUTHOR INFORMATION**

E-mail addresses for correspondence: gcolherinhas@gmail.com (G.C); fileti@gmail.com (E.E.F.), vvchaban@gmail.com (V.V.C.)